\documentclass[shortnote]{jpsj2} 
\title{Distribution of Microscopic Energy Flux in Equilibrium State}

\author{
Takashi \textsc{Shimada}\thanks{E-mail address: shimada@ap.t.u-tokyo.ac.jp},
Fumiko \textsc{Ogushi},\\
and Nobuyasu \textsc{Ito}
}

\inst{
Department of Applied Physics, School of Engineering,\\
The University of Tokyo,\\
Hongo, Bunkyo-ku, Tokyo 113-8656, Japan
}

\abst{
The distribution function $P(j)$ of the microscopic energy flux, $j$,
in equilibrium state is studied.
It is observed that $P(j)$ has a broad peak in small $j$ regime
and a stretched-exponential decay for large $j$.
The peak structure originates in a 
potential advection term and energy transfer term between the particles.
The stretched exponential tail comes from the momentum energy advection term.
}

\kword{transport, microscopic flux, heat conduction, Lennard-Jones}

\begin{document}
\maketitle

Recent studies on reproducing heat conduction in Hamilton systems
revealed that anomalous thermal conductivity is ubiquitous
in wide variety of systems even if those have strong mixing property,
especially in those the total momentum is conserved\cite{Lepri,Shimada,Murakami,Shiba}.
What is typical in those systems is that
the {\it thermal conductivity} slowly diverges with system size
while the temperature profile has well-scaled normal form.
Those facts indicate that there are both
diffusive and ballistic transport processes in such systems.
Microscopic description and understanding for such non-diffusive energy transport,
beyond its mere evaluation from the anomaly of the macroscopic transport coefficient,
is expected so that it will be controlled and designed.
The distribution of microscopic energy flux will be the most elementary quantity for this purpose.
However, it is not clear what will be the {\it normal} distribution for
microscopic energy flux since the energy flux has no direct relation with conserved quantities.
In this note we discuss the distribution of microscopic energy flux
in thermal equilibrium state.

Let us consider the local energy flux \mbox{\boldmath $J$}
at a small volume element $V$ in particle systems in equilibrium state. 
Microscopic expression of \mbox{\boldmath $J$} is
$
\mbox{\boldmath $J$} = \frac{1}{V}\sum_i \mbox{\boldmath $j$}_i,
$
where $\mbox{\boldmath $j$}_i$ denotes the energy flux
carried by each particle in the volume element $V$,
hence the summation is taken for the particles in the volume.
In the following we will focus on the ensemble distribution of $j$.

For $d-$dimensional ideal gas, this local energy flux $J$
is simply an average of single-particle energy fluxes and
the each single-particle flux $j$ is expressed as
\begin{equation}
\mbox{\boldmath $j$} = e\cdot \mbox{\boldmath $v$}
= e \cdot \frac{\mbox{\boldmath $p$}}{m}
= \frac{p^2 \mbox{\boldmath $p$}}{2m^2},
\label{momentumj} 
\end{equation}
where $e$, \mbox{\boldmath $v$}, \mbox{\boldmath $p$} and $m$
denote kinetic energy, velocity, momentum and mass of each particle, 
respectively. 
Since equilibrium distribution of momentum is 
\begin{equation}
P(\mbox{\boldmath $p$}) \mbox{d}\mbox{\boldmath $p$}
=
N
\exp\left\{\frac{-\beta p^2}{2m}\right\} \mbox{d}\mbox{\boldmath $p$}
\quad
\left( \beta = (k_{\rm B}T)^{-1} \right),
\label{pmomentum}
\end{equation}
where $k_{\rm B}$, $T$, and $N$
denote Boltzmann's constant, temperature, and a normalization factor, respectively,
and the Jacobian for the transformation from
\mbox{\boldmath $p$} to \mbox{\boldmath $j$}
is calculated as
\begin{equation}
\left| \frac{d \mbox{\boldmath $j$}}{d \mbox{\boldmath $p$}} \right|
= 3 \left( \frac{p^2}{2m^2} \right)^d
= 3 \left( \frac{j^2}{2m^2} \right)^\frac{d}{3},
\label{jacobian}
\end{equation}
the equilibrium distribution of energy flux is obtained as
\begin{eqnarray}
P(\mbox{\boldmath $p$})\mbox{d}\mbox{\boldmath $p$}
&=& P(\mbox{\boldmath $p$})
\frac{\mbox{d} \mbox{\boldmath $p$}}{\mbox{d} \mbox{\boldmath $j$}}
\mbox{d} \mbox{\boldmath $j$}\\
&=& \frac{N}{3}
\left( \frac{2m^2}{j^2} \right)^\frac{d}{3}
\exp\left\{-\left(\frac{m}{2}\right)^\frac{1}{3} \beta j^\frac{2}{3}\right\}
\mbox{d} \mbox{\boldmath $j$}.
\end{eqnarray}
Therefore
\begin{equation}
P(j) = N'
j^\frac{d-3}{3}
\exp\left\{- \left( \frac{m}{2} \right)^\frac{1}{3} \beta \, j^\frac{2}{3}\right\},
\label{pjk}
\end{equation}
where $N'$ is a normalization factor which does not depend on $j$.
It is notable that the microscopic energy flux distribution
in equilibrium state
has a stretched exponential form,
although it is just a transcription of Boltzmanian distribution function.

For real gas with interaction between particles,
the local energy flux $J$ may not be determined
solely from the distribution of single particle flux $\mbox{\boldmath $j$}_i$
since the correlation among them may be relevant.
Yet the single-particle distribution
is still essential and is a good observable for simulational physics.
For the systems which have a Hamiltonian of
\begin{equation}
{\cal H} = \sum_i \frac{p_i^2}{2m^2} +
\sum_{i < j} U_{ij}(\mbox{\boldmath $q$}_i,\mbox{\boldmath $q$}_j),
\end{equation}
the energy flux transported by a particle $i$ can be expressed as
\begin{equation}
\mbox{\boldmath $j$}_i
=
\frac{p_i^2 
\mbox{\boldmath $p$}_i}{2m^2}
+ \sum_j \left\{
\frac{ U_{ij} \mbox{\boldmath $p$}_i}{2m}
- \left( \mbox{\boldmath $q$}_i - \mbox{\boldmath $q$}_j \right)
\left(
\frac{\partial U_{ij}}{\partial \mbox{\boldmath $q$}_i} \cdot \frac{\mbox{\boldmath $p$}_i}{m}
\right) \right\},
\end{equation}
where the first term is the same term as eq. (\ref{momentumj}).
The second and third terms,
come from the interaction,
denote the advection term of the local potential energy
and the energy transfer via the potential force, respectively.
In the following, we will denote those three flux terms as
$\mbox{\boldmath $j$} =
\mbox{\boldmath $j$}_K + \mbox{\boldmath $j$}_U + \mbox{\boldmath $j$}_F$
with
$\mbox{\boldmath $j$}_K = \frac{p_i^2}{2m^2} \, \mbox{\boldmath $p$}_i$,
$\mbox{\boldmath $j$}_U = \sum_j \frac{ U_{ij} \mbox{\boldmath $p$}_i}{2m}$,
and
$\mbox{\boldmath $j$}_F =
- \sum_j \left( \mbox{\boldmath $q$}_i - \mbox{\boldmath $q$}_j \right)
\left(
\frac{\partial U_{ij}}{\partial \mbox{\boldmath $q$}_i} \cdot \frac{\mbox{\boldmath $p$}_i}{m}
\right)$.
\begin{figure}[thbp]
\rotatebox{-90}{\includegraphics[width=6.0cm]{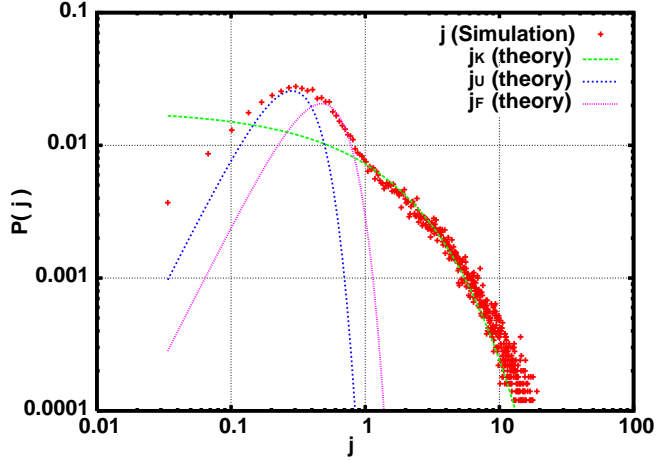}}
\caption{
Distributions of microscopic energy flux.
Points:
$P(j)$ obtained in Lenard-Jones particle system
with $\epsilon \beta = 1/3.4$, and $\rho=0.25$.
Total number of particles is $843$. 
Lines:
theoretical predictions from the equation (\ref{pjk}) and (\ref{p1st}).
}
\label{ogushiLJ}
\end{figure}
\begin{figure}[thbp]
\rotatebox{-90}{\includegraphics[width=6.0cm]{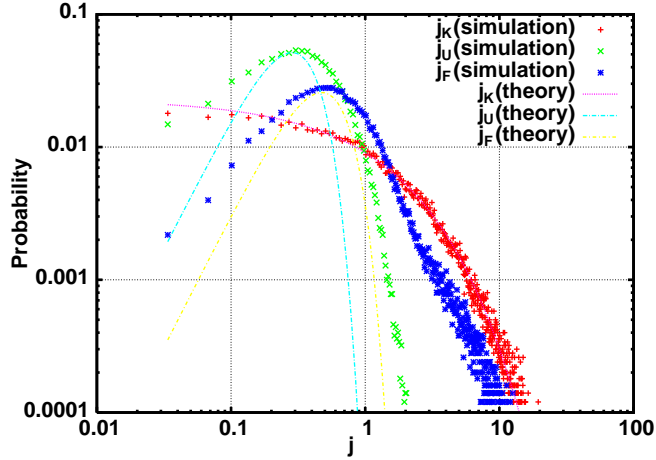}}
\caption{
Distributions of $J_K$, $J_U$, and $J_F$.
Points:
simulation results obtained from the same system as Fig\ref{ogushiLJ}.
Lines:
theoretical predictions.
}
\label{fits}
\end{figure}

Since there is no correlation between the distributions of momenta
and coordinates in the equilibrium ensemble,
the expectation value of the flux for a given momentum
{\boldmath $p$} can be written as
\begin{equation}
<\mbox{\boldmath $j$}(\mbox{\boldmath $p$})>_q
=
\frac{p^2 \mbox{\boldmath $p$}}{2m^2}
+ \Psi (\beta) \cdot \mbox{\boldmath $p$}
\equiv \mbox{\boldmath $j$}_K + \tilde{\mbox{\boldmath $\jmath$}},
\label{jkju}
\end{equation}
where $<>_q$ denotes the ensemble average over coordinate space,
$\Psi(\beta)$ is a tensor calculated as
\begin{equation}
\Psi(\beta)
= \left< \frac{1}{m} \sum_j \left(
\frac{U_{ij}}{2}
- (\mbox{\boldmath $q$}_i-\mbox{\boldmath $q$}_j)
\frac{\partial U_{ij}}{\partial \mbox{\boldmath $q$}_i}
\right) \right>_q,
\end{equation}
and
$\tilde{\mbox{\boldmath $\jmath$}} =
<\mbox{\boldmath $j$}_U + \mbox{\boldmath $j$}_F>_q$.
Note that $\Psi(\beta)$ is essentially the virial.

The equilibrium distribution of $\tilde{\jmath}$ is calculated as
\begin{equation}
P(\tilde{\jmath})
=
N''
|\Psi(\beta)|^{-d} \tilde{\jmath}^{d-1}
\exp\left\{- \frac{\beta}{2m^2|\Psi(\beta)|^2} \, \tilde{\jmath}^2\right\},
\label{p1st}
\end{equation}
with a normalization factor $N''$.
This $\tilde{\jmath}$,
which stems from the potential and has $p$-first order form,
becomes relevant for small $p$, and hence for small $j$.
Since the flux $\mbox{\boldmath $j$}_U + \mbox{\boldmath $j$}_F$
is a product of $p$ and potential-related term,
the width of the distribution of itself
is expected to be broader than the one of $\tilde{\jmath}$
and they are related by
\begin{equation}
\log(\sigma_{j_U + j_F})
=
\log(\sigma_{\Psi(\beta)}) + \log(\sigma_{\tilde{\jmath}}).
\label{width}
\end{equation}

Eqs.(\ref{pjk}), (\ref{jkju}), and (\ref{p1st})
imply that the distribution of the total local energy flux $P(j)$
shows crossover from Arrhenius-type region,
in which potential term is dominant,
to non-Arrhenius-type region,
in which momentum energy advection is dominant.

We now examine the argument using three dimensional
Lenard-Jones particle system with the potential
\begin{equation}
U_{ij} =
4 \epsilon \left\{
\left( \frac{1}{r_{ij}} \right)^{12}
-
\left( \frac{1}{r_{ij}} \right)^{6}
\right\}
\,\,\,\,
\left( r_{ij} = |\mbox{\boldmath $q$}_i - \mbox{\boldmath $q$}_j| \right)
\label{LJpotential}
\end{equation}
and the uniform mass ($m = 1$).
This system is known to have normal thermal conductivity\cite{Ogushi}.
The distribution $P(j)$ is estimated by computer simulation
at high temperature ($\epsilon \beta = 1/3.4$)
and not dense (number density $\rho=0.25$) condition (Fig.\ref{ogushiLJ}).
We can see that the distribution with crossover
discussed in this note appears in the simulation result.
Since the density is low and the temperature is high,
it is expected that $\Psi(\beta)$ can be well estimated
by taking the average for single pair using two-body distribution.
The lines in Fig.\ref{ogushiLJ} show the distributions calculated
using the analytic result for the second virial coefficient\cite{LJZ}.
As shown in Fig.\ref{ogushiLJ}, the envelope of
theoretical predictions for $P(j_K)$, $P(j_U)$, and $P(j_F)$
well explains the profile of simulational result of $P(j)$
i.e. the peak position and the stretched exponential tail.
It is also confirmed that
theoretical prediction and simulational result
show good agreement in each $P(j_K)$, $P(j_U)$, and $P(j_F)$,
except for large $j_F$ regime where the latter has
stretched exponential tail
which originates from high-energy collision events (Fig.\ref{fits}).

We have focused directly on the distribution of the microscopic energy flux.
From the microscopic point of view,
there are many ways to transport the energy
without average momentum flow.
Asymmetry in the higher order cumulant of $P(p)$
is a possible candidate,
although it is known to be usually hard
to detect the deviation directly in the distribution of dynamical variable.
Correlations between momentum variables and coordinate variables
will be an another possibility.
For local heat flux,
spatial correlations among particles is also essential
for anomalous behaviors and that should be detected
as the deviation from the independent sum of the single particle distribution.
Furthermore,
to consider the boundary in non-equilibrium states,
for instance\cite{Ogushi2},
anomaly can be estimated only from microscopic quantities.
The argument we took and the obtained {\it normal} distribution
give good basis for considering those problems.


\begin{thebibliography}{9}
\bibitem{Lepri}
S. Lepri, R. Livi, and A. Politi,
Phys. Rev. Lett. \textbf{78} (1997) 1896.,
and 
Physics Reports \textbf{377} (2003) 1.,

\bibitem{Shimada}
T. Shimada, T. Murakami, S. Yukawa, K. Saito, and N. Ito,
J. Phys. Soc. Jpn. \textbf{69} (2000) 3150.

\bibitem{Murakami}
T. Murakami, T. Shimada, S. Yukawa, and N. Ito,
J. Phys. Soc. Jpn. \textbf{72} (2003) 1049.

\bibitem{Shiba}
H. Shiba, S. Yukawa, and N. Ito,
J. Phys. Soc. Jpn. \textbf{75} (2006) 103001.

\bibitem{Ogushi}
F. Ogushi, S. Yukawa, and N. Ito,
J. Phys. Soc. Jpn. \textbf{74} (2005) 827.

\bibitem{LJZ}
P. Vargas, E. Mu\~{n}oz, and L. Rodriguez,
Physica A \textbf{ 290} (2001) 92.

\bibitem{Ogushi2}
F. Ogushi, S. Yukawa, and N. Ito,
J. Phys. Soc. Jpn. \textbf{75} (2006) 073001.
\end{thebibliography}
\end{document}